# Imaging Dynamic Microtubules and Associated Proteins by Simultaneous Interference-Reflection and Total-Internal-Reflection-Fluorescence Microscopy

Yazgan Tuna, Amer Al-Hiyasat and Jonathon Howard

*Department of Molecular Biophysics & Biochemistry and Department of Physics, Yale University, New Haven, 06511 Connecticut, USA*

Several techniques have been employed for the direct visualization of cytoskeletal filaments and their associated proteins. Total-internal-reflection-fluorescence (TIRF) microscopy has a high signal-to-background ratio, but it suffers from photobleaching and photodamage of the fluorescent proteins. Label-free techniques such as interference reflection microscopy (IRM) and interference scattering microscopy (iSCAT) circumvent the problem of photobleaching but cannot readily visualize single molecules. Here, we present a protocol for combining IRM with a commercial TIRF microscope for the simultaneous imaging of microtubule-associated proteins (MAPs) and dynamic microtubules *in vitro*. Our protocol allows for high-speed observation of MAPs interacting with dynamic microtubules. This improves on existing two-color TIRF setups by eliminating both the need for microtubule labeling and the need for several additional optical components, such as a second excitation laser. We image both channels on the same camera chip to avoid image-registration and frame-synchronization problems. We demonstrate our setup by visualizing single kinesin molecules walking on dynamic microtubules.

## INTRODUCTION

Total Internal Reflection Fluorescence (TIRF) microscopy is commonly employed for the visualization of single fluorescent molecules. In comparison to epifluorescence imaging, TIRF achieves superior background suppression, allowing for the high-resolution localization and tracking of single fluorophores. For this reason, TIRF is the preferred method for visualizing fluorescently labeled microtubule-associated proteins and is frequently used to image microtubules[1,2].

To investigate the regulation of microtubule dynamics by MAPs, it is often necessary to image both microtubules and MAPs simultaneously. Most existing methods for this purpose are expensive or suffer from technical drawbacks. Simultaneous two-color TIRF, for instance, requires two excitation lasers and two cameras. In addition to the high cost, the need for separate cameras poses frame-synchronization and image-registration problems. This need can be circumvented if a rotating filter cube is used to physically switch between excitation lasers in consecutive frames[3]. In such a setup, a single camera chip can be used, and the frames alternate between images of microtubules and of MAPs. This technique, however, is limited by the speed of the filter change, which typically restricts the frame rate to less 0.5 frames per second[3]. Such a frame rate is insufficient to resolve fast dynamic processes, such as the shrinkage of a microtubule, which occurs at a velocity up to 500 nm/s, the walking of a kinesin, with a velocity on the order of 800 nm/s, or the diffusion of a MAP, which can occur with diffusion coefficients exceeding 0.3 μm$^2$/s [4]. This is particularly problematic when tracking the relative positions of two moving targets in each channel, such as the position of a MAP relative to the position of a moving microtubule tip[5].

In addition to these optical constraints, two-color TIRF microscopy requires MAPs and microtubules to be labeled with different fluorophores whose emission spectra are sufficiently separated. Fluorescent labeling of tubulin can alter microtubule dynamics[6], and the photobleaching of fluorophores limits the imaging speed[7]. Because of these issues, label-free imaging techniques have been developed to visualize microtubules. These include interferometric scattering microscopy (iSCAT)[8, 9], rotating-coherent-scattering microscopy (ROCS)[10], spatial light interference microscopy (SLIM)[11], and interference reflection microscopy (IRM)[12, 13]. These techniques allow fast label-free imaging of microtubules without the drawbacks of fluorescence imaging, but they cannot be used to visualize single MAPs.

Of these label-free techniques, IRM stands out for its low cost and its modest demands on instrumentation. We have recently presented a protocol for combining IRM with a commercial TIRF microscope, allowing microtubules and fluorescent MAPs to be imaged in alternating frames[3, 13]. Here, we present a protocol for modifying this setup to simultaneously capture TIRF and IRM images on a single camera chip. This involves the addition of an inexpensive beam splitter in the excitation path to simultaneously illuminate the sample with a TIRF laser and IRM LED light source. A modified commercial image splitter is also used to spectrally separate the TIRF and IRM signals and project them on separate halves of the same camera chip. Our protocol describes how this setup can be used to image dynamic microtubules and MAPs. We demonstrate the capability of our apparatus by presenting the first visualization of kinesin-1 proteins walking on shrinking microtubules, which is captured at a frame rate of 10 s$^{-1}$.

**PROTOCOL**

**1. Preparation of flow chambers**

We employ a microfluidic system consisting of 5-Polydimethylsiloxane (PDMS) microchannels fixed on a functionalized cover glass. This allows the rapid exchange of reagents during imaging. For kinesin motility assays, we used piranha-cleaned silanized cover glasses prepared as described in Gell et. al. (2010)[1]. A simpler cleaning procedure (easy-clean) is to sonicate cover glasses in isopropanol followed by methanol for three cycles each 20 minutes long. The appropriate cover glass surface chemistry will depend on the specific assay being imaged but must withstand 80 °C temperatures for the PDMS microchannels to adhere.

    **1.1. Preparation of PDMS channels**
        1.1.1. To prepare a master mold, cut strips of single-sided stationary tape to the desired flow channel size. Adhere the strips to the bottom of a 10 cm petri dish, arranging them side to side with at least 1 cm spacing between strips. NOTE: If precise channel dimensions are required, the mold can be fabricated on silicon wafers using photolithography.
        1.1.2. To prepare the PDMS polymer, combine the curing agent and the base elastomer in a 1:10 mass ratio. Mix for 2 minutes.

NOTE: This mixing ratio can be varied to tune the stiffness of the polymer. A higher ratio of base to curing agent results in a softer polymer.
1.1.3. Degas the mixture in a vacuum chamber until all bubbles disappear.
1.1.4. Pour the mixture onto the master mold in a 0.5 cm thick layer, taking care to avoid creating bubbles.
1.1.5. Bake the mixture in a preheated oven at 70 °C for 40 minutes.
NOTE: If the PDMS block is thick (>1cm), a longer curing time may be required. Continue heating in 5-minute increments until it is fully cured.
1.1.6. Cut out the structured regions of the polymer. Punch holes at each end of the channel using a PDMS puncher.
NOTE: PDMS channels can be stored in dry environments for extended periods but should be cleaned prior to use.
1.1.7. Clean the structured side of the PDMS block. Use stationary tape to remove large particles. Rinse with isopropanol and then methanol. Repeat these rinses three times, then rinse with milli-Q water and blow-dry the surface.
1.1.8. Plasma clean the PDMS using oxygen or air plasma.
1.1.9. Place the plasma cleaned PDMS on an appropriately cleaned cover glass and heat on a hot plate at 80°C for 15 minutes.
NOTE: Epoxy resin can be applied to the sides of the PDMS block to better adhere it to the cover glass
1.1.10. Insert appropriately sized LDPE tubing into the holes. Connect the outlet tubing to a 0.5 ml syringe.
NOTE: We used tubing with an inner diameter of 0.023 inches connected to the PDMS via a 0.025 inch outer diameter metal adaptor.
1.1.11. Solutions can be flowed into the microchannels by immersing the inlet tube in the solution and drawing the required volume with the syringe.

2. **Optical setup**
    2.1. **Modification of the microscope (Figure 1)**
        2.1.1. Following Mahamdeh and Howard. 2019[13], modify a TIRF microscope to enable IRM imaging. Replace the 50/50 (Reflectance (R)/Transmission (T)) beam splitter used in the filter wheel of the microscope with a 10/90 (R/T) beam splitter.
        2.1.2. Insert an image splitter between the camera and the microscope.
        NOTE: In this protocol, we used a Photometrics DualView Lambda (now OptoSplit II) image splitter.
        2.1.3. In the filter cube of the image splitter: Insert a 10/90 (R/T) beam splitter. Place a 600nm long-pass filter in front of the reflected beam, and an appropriate fluorescence emission filter in front of the transmitted beam.
        NOTE: The R/T ratio can be changed to tune what fractions of the IRM and TIRF emission light are collected.
        2.1.4. Align the image splitter according to the manufacturer's specifications. This will spatially separate the TIRF and IRM signals on the camera chip.
3. **Imaging dynamic microtubules and single Kinesin molecules**
    3.1. **Surface functionalization and passivation**
        3.1.1. Into the reaction chamber, flow BRB80 buffer (80 mM Na-PIPES, 1 mM EGTA, 1 mM $MgCl_2$, titrated to pH 7.8 with NaOH).

- 3.1.2. Flow anti-biotin solution diluted to 0.025 mg/mL in BRB80 and incubate at room temperature for 10 minutes.
- 3.1.3. Wash the channel with BRB80.
- 3.1.4. Flow in F-127 solution (1% pluronic F127 (w/v) dissolved in BRB80 overnight) and incubate for at least 20 minutes for surface passivation.
  NOTE: If easy-cleaned cover glasses are used instead of silanized cover glasses, passivate using 2 mg/mL casein in BRB80 for >20 min at room temperature.
- 3.1.5. Wash the channel with BRB80.

**3.2. Preparation for imaging**
- 3.2.1. If temperature control is required, use an objective heater and set it to the correct temperature.
  NOTE: In this protocol, we performed all the experiments at 28 °C.
- 3.2.2. Place the sample on the microscope stage and turn on the epi-illumination light source for IRM imaging.
  NOTE: In this protocol, we used a Lumencor Sola LED light source.
- 3.2.3. Focus the microscope on the sample surface. Finding the correct focal plane is easiest near the PDMS-solution interface. In IRM, the aqueous interior of the channel should appear much brighter than the PDMS polymer.
- 3.2.4. Pick a field of view near the center of the channel.

**3.3. Growing biotinylated GMPCPP-stabilized microtubule seeds.**
- 3.3.1. In a 0.6 μl centrifuge tube, prepare a 50 μl solution containing 1 mM GMPCPP, 1 mM $MgCl_2$, and 2 μM biotinylated tubulin (5-10% labeling stoichiometry) in BRB80.
  NOTE: The correct labeling stoichiometry can be achieved by combining high-density biotinylated tubulin with unlabeled bovine tubulin (commercially available from e.g. Cytoskeleton Inc.) in the correct ratio.
- 3.3.2. Incubate the solution on ice for 5 min, then incubate at 37 °C for 12.5 min.
  NOTE: The length of the seeds can be controlled by tuning the polymerization time.
- 3.3.3. Stop the polymerization by adding 100 μl room temperature BRB80.
- 3.3.4. Spin down the solution in an ultracentrifuge at room temperature (126,000 g, 5 min). Discard the supernatant to remove unpolymerized tubulin.
  NOTE: we used a table-top Beckman-Coulter air-driven ultracentrifuge.
- 3.3.5. Add 200 μl room temperature BRB80 to the pellet. Resuspend the pellet by pipetting gently but thoroughly. Use a 200 μl pipette with a cut tip to reduce the shearing of the microtubules.

**3.4. Growing dynamic GDP "extensions"**
The seeds will be immobilized on the anti-biotin surface of the flow channel. Dynamic GTP/GDP "extensions" will be grown from the ends of the immobilized seeds.
- 3.4.1. Dilute the seeds 20X in BRB80. Flow the diluted seeds into the reaction chamber.
- 3.4.2. Monitor the reaction via IRM. Seeds will gradually "land" onto the surface and bind to it. When the desired density of seeds is achieved, wash out the excess with BRB80.

- 3.4.3. Prepare a solution of 0.1 µm 200-fold diluted TetraSpeck Microspheres in BRB80.
- 3.4.4. Prepare a microtubule extension mixture: 12 µM unlabeled tubulin, 1 mM GTP, 5 mM Dithiothreitol (DTT) in BRB80 buffer.
- 3.4.5. Flow in at least one channel volume of the extension mixture. The reaction temperature should be 28 °C.
- 3.4.6. Microtubule extensions will grow from the seeds over time. The equilibrium length is usually achieved in less than 20 min.
- 3.4.7. The dynamic microtubules are now ready for imaging. Fluorescent MAPs can be added and visualized on the microtubules.

**3.5. Kinesin motility assay**

Here, we describe a protocol for visualizing motile GFP-labelled kinesin-1 on shrinking microtubules. We expressed and purified a truncated rat kinesin-1 construct fused to eGFP (rKin430-eGFP) as described in Rogers et al. (2001)[14] and Leduc et al. (2007)[15].

- 3.5.1. Prepare motility buffer: 1 mM ATP and 0.2 mg/mL casein in BRB80.
- 3.5.2. Dilute kinesin-eGFP in motility buffer to 10 nM.
- 3.5.3. Prepare a 2X oxygen scavenger solution to counteract oxidative photobleaching (80 mM glucose, 80 mg/mL glucose oxidase, 32 mg/mL catalase, 0.2 mg/mL casein, 20 mM DTT) supplemented with 2 mM ATP.
- 3.5.4. Combine 10 parts oxygen scavenger, 9 parts BRB80, and 1 part 10 nM kinesin-EGFP for a final kinesin concentration of 0.5 nM.
  NOTE: The total volume should be at least 1.5-fold larger than the channel volume.
- 3.5.5. Set the imaging settings on the microscope software.
  NOTE: In this protocol, we recorded at 10 fps with 100 ms exposure time.
- 3.5.6. Begin imaging and flow the kinesin solution into the chamber.
- 3.5.7. After the measurement is complete, record a short (about 5 seconds) video in which the stage is slowly translated in a circular or lateral motion. The median projection of this video will serve as a background image and will be subtracted from the raw measurements.

4. **Image processing and analysis**

Image processing was carried out using NIH ImageJ2 (imagej.nih.gov/ij/). We developed a macro to automate splitting and alignment of the TIRF and IRM channels. This macro requires the GaussFit_OnSpot plugin to be installed (available on the ImageJ plugins repository).

- **4.1.** From the background recording, create a median projection in ImageJ using Image > Stacks > Z-project.
- **4.2.** Subtract the median background projection from the raw image data via Process > Image Calculator.
- **4.3.** For image registration, pick a collection of TetraSpeck beads that are near the microtubule of interest. These beads will be used to align the TIRF and IRM images.
- **4.4.** For each bead in this collection, use the multipoint selection tool to mark the approximate location in the TIRF channel, and then the corresponding location in the IRM channel.

NOTE: For example, if there are two beads (1 and 2), the multipoint selection would have four points in the following order: (1) bead 1 in TIRF, (2) bead 1 in IRM, (3) bead 2 in TIRF, (4) bead 2 in IRM.

**4.5.** Run the provided ImageJ macro (ImageSplitterRegistration.ijm). The macro performs the following steps:
  4.5.1. Estimate the center location of each bead by fitting to a Gaussian.
  4.5.2. For each bead, compute the displacement vector separating the center in the TIRF channel from the center in the IRM channel
  4.5.3. Average this displacement vector across all beads.
  4.5.4. Split the TIRF and IRM channels into separate images.
  4.5.5. Translate the TIRF image by the average displacement vector computed in 4.4.1.
  4.5.6. Overlay the TIRF and IRM images in a multi-channel .tiff file.

**RESULTS**

The optical setup is schematized in Figure 1. Both IRM illumination and TIRF excitation light are directed to the back aperture of the objective via a 10/90 (R/T) beam splitter (BS1). The emitted signal passes through the same beam splitter (BS1) and is reflected to the image splitter via a mirror (M1). The components of the image splitter (enclosed with dashed lines in Figure 1) separate the IRM and TIRF signals via a 90/10 (R/T) beam splitter together with appropriate spectral filters. Finally, the split images are projected onto the camera ship for visualization.

In a well-aligned microscope, the camera image should display a halfway split image as presented in Figure 2. Surface-bound microtubules should be easily visible in the IRM channel[13], and fluorescent kinesin should be visible in the TIRF channel.

The TetraSpeck beads that we used to align and register the two channels appear as bright spots in the TIRF images and dark spots in the IRM images. Although the beads are visible in the raw data, background subtraction improves the contrast significantly (see Figure 2). The background image used for the subtraction is the temporal median of a video that is recorded with a moving stage. As described in the protocol, image alignment was performed by selecting a collection of beads near the region of interest and executing the macro we provided (imageSplitterRegistration.ijm). The macro fits the points to Gaussians and aligns the images by minimizing the average distance between the center points of the fits in each channel. This process is represented in Figure 2, which shows the good alignment of the TetraSpeck beads (green in the TIRF channel, black in the IRM channel).

Finally, we present the capabilities of our simultaneous imaging setup by observing single kinesin molecules walking towards the shrinking ends of microtubules. In Figure 3, we show a kymograph of eGFP-labeled kinesin molecules (green) walking on a shrinking microtubule (gray). We also present a series of snapshots from the recording from which the kymograph was generated.

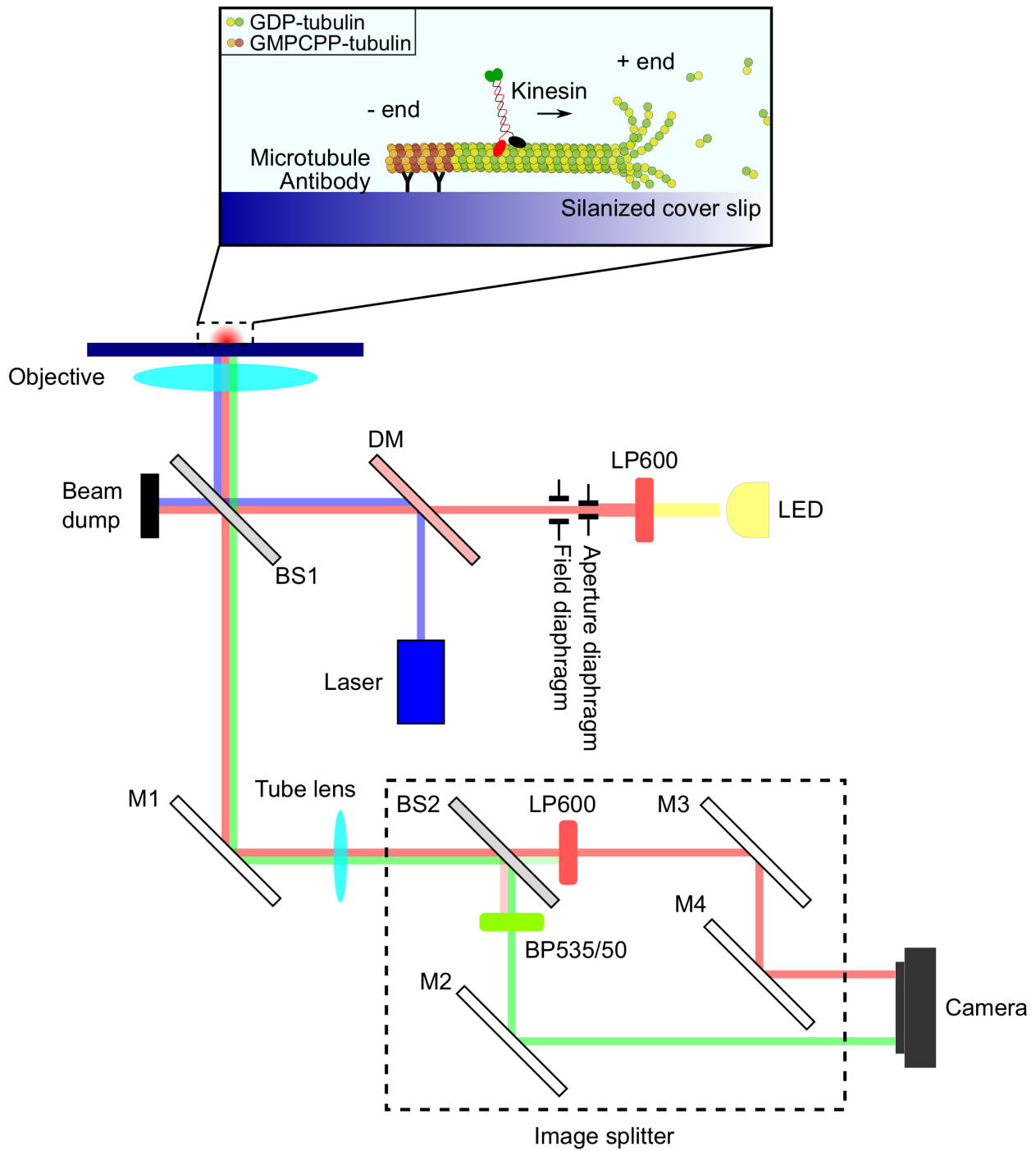

**Figure 1. Schematic representation of the optical setup for simultaneous IRM and TIRF imaging of kinesin motility.** Epi-illumination from an LED light source passes through the aperture diaphragm and reaches the 10/90 (R/T) beam splitter (BS1). The beam splitter partially reflects the red IRM illumination light and the 488nm TIRF excitation light up to the objective to illuminate the sample. The signal from the sample is collected by the same objective and directed to the image splitting assembly where IRM and TIRF images are spectrally and spatially separated before reaching the camera chip. Abbreviations: R/T: Reflected/Transmitted: LP600: Long pass filter (591 nm), DM: Dichroic mirror; BS1 and BS2: beam splitter 1 and 2; M1, M2, M3, M4: Mirrors.

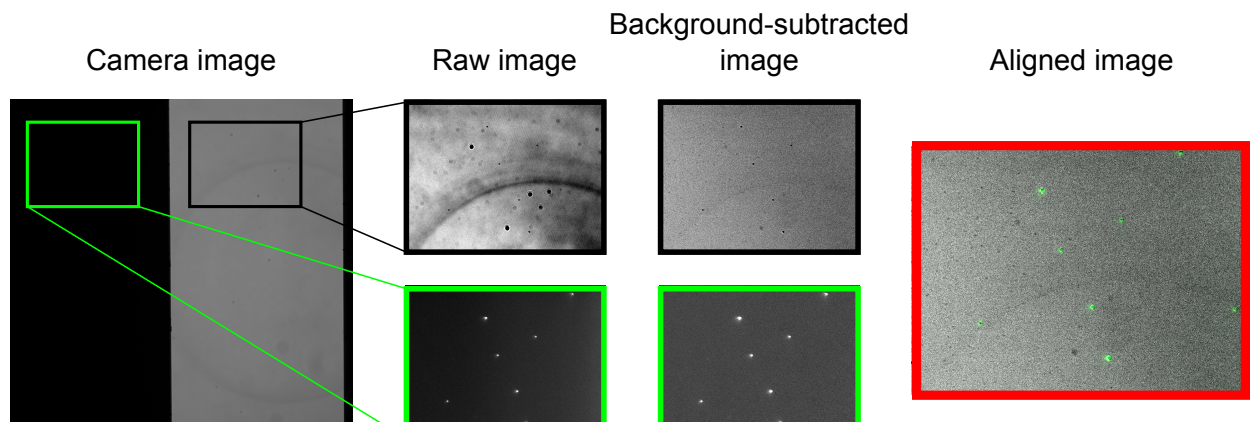

**Figure 2. Background subtraction and image alignment.** The TIRF and IRM images appear simultaneously on two halves of the same camera chip (left). Temporal median background subtraction increases the contrast of the beads (background-subtracted image). IRM and TIRF images are aligned by translation based on the localization of selected beads (right).

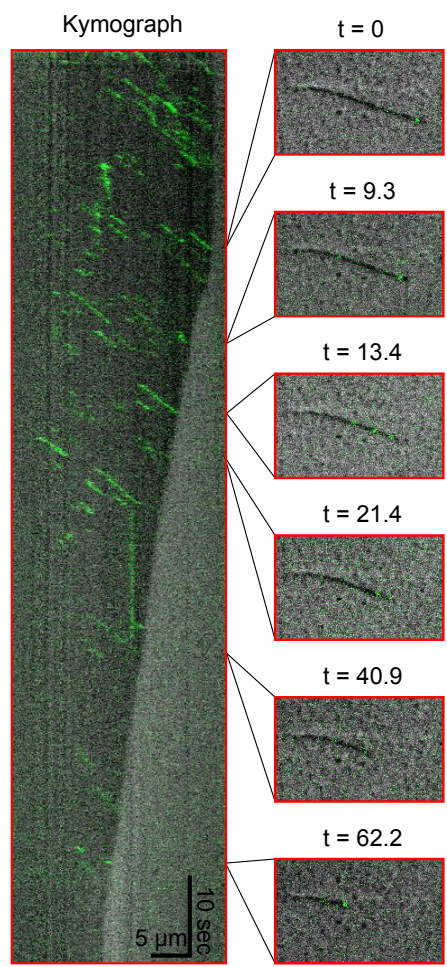

**Figure 3. Kymograph and snapshots of Kinesins' movement during microtubule shrinkage.** The kymograph (left) shows eGFP-kinesin-1 (green) walking towards the plus end of the microtubule (dark grey). Images from the time series from which this kymograph was generated are shown (right).

**DISCUSSION:**

Studying the regulation of microtubule dynamics by microtubule-associated proteins (MAPs) often requires simultaneous imaging of microtubules and MAPs. Fluorescence microscopy techniques such as TIRF are often employed for this purpose but are limited by the drawbacks of fluorescence imaging, which include photobleaching, photodamage, and the need for fluorophore labeling. Label-free methods, such as IRM, are suitable for visualizing microtubules but are not capable of imaging single fluorophores. In this protocol, we combine label-free IRM imaging and TIRF microscopy for simultaneous imaging of dynamic microtubules and MAPs.

The IRM setup employs an LED illumination source filtered to >600nm, while the TIRF setup utilizes a 488nm laser. We used an inexpensive plate beam splitter to both reflect the illumination light onto the sample and transmit the collected signal to the detector (Figure 1). To minimize the loss of the single-molecule signal, we chose a beam splitter with 10% reflectance and 90% transmission. The 90% loss in the illumination light intensity is compensated for by increasing the power of the illumination laser and LED.

Spectral separation of the signals was achieved using a 90/10 (R/T) beam splitter and two spectral filters (Long-pass 600 nm for IRM and band-pass 535/50 nm for TIRF). The spectrally separated IRM and TIRF signals are projected onto two halves of a single camera chip using a commercial Photometrics DualView Lambda image splitter assembly. The use of a 90/10 beam splitter sacrifices 90% of the IRM signal, but this is compensated for by increasing the intensity of the LED illumination source. A dichroic mirror could also be used here to more efficiently separate the IRM and TIRF signals. We included fluorescent TetraSpeck microspheres in our assays. These beads enable accurate alignment of the TIRF and IRM images and serve as a reference for focusing the objective.

The most critical optical element in this protocol is the high numerical aperture (NA) objective. This is essential not only to achieve total internal reflection but also to maximize the collection efficiency and image contrast. The quality of the obtained images also depends on the cleanliness of the glass surface and on the acquisition of a clear background image to correct for nonuniform illumination and remove static features. For IRM imaging, we recommend using long-wavelength illumination (>600 nm) to minimize the photodamage of microtubules and proteins. This is particularly important if a white LED light source is used, in which case a long-pass filter should be included to remove any UV light.

In conclusion, our protocol implements the simultaneous imaging of dynamic microtubules and MAPs via combined IRM and TIRF microscopy. This technique allows for label-free high-speed imaging of dynamic microtubules and simultaneous high-resolution visualization of fluorescent MAPs. In comparison to the filter-cube-switching technique, in which are alternated between capturing images of microtubules and MAPs, our setup is capable of much higher frame rates because it does not depend on the physical rotation of a filter cube. The speed of our setup is limited only by the bleaching times of the fluorophores on the MAPs. In comparison to two-color TIRF imaging techniques, our technique employs a less demanding optical setup. Furthermore, it circumvents the need for fluorophore labeling of microtubules.

To demonstrate the capabilities of this technique, we simultaneously visualized two fast dynamical processes via IRM and TIRF: the shrinkage of a microtubule and the walking of a fluorescent kinesin molecule. We have also previously employed this technique to visualize the fast diffusion of spastin on shrinking microtubules[5]. Beyond this application to MAPs and microtubules, our protocol can be used to visualize single fluorescent molecules simultaneously with any macromolecular structure massive enough to be visualized via IRM, such as a cell membrane or actin filament.

**TABLE OF MATERIALS**

| Name of Material/ Equipment | Company | Catalog Number |
|---|---|---|
| Microscope | Nikon | Ti-Eclipse |
| Nikon Plan Fluor 100X/0.5-1.3 Iris objective | Nikon | MRH02902 |
| Microscope objective heater | okolab | H401-T-DUAL-BL |
| LED light source | Lumencor | Lumencor sola light engine |
| 10/90 (R/T) beam splitter | Thorlabs | BSN10R |
| 90/10 (R/T) beam splitter | Thorlabs | BSX10R |
| Long-pass filters | Thorlabs | FELH0600 |
| Band-pass filter | Newport | HPM535-50 |
| Image splitter | Teledyne-Photometrics Imaging | OptoSplit II |
| Zyla 4.2 camera | Andor | Zyla 4.2 |
| PDMS and curing agent | Electron Microscopy Sciences | Sylgard 184 (24236-10) |
| PDMS puncher | World Precision Instruments LLC | 504529 |
| Plasma cleaner | Harrick Plasma | DPC-32G |
| Table-top ultracentrifuge | Beckman Coulter | 340400 |
| Anti-biotin antibody | Sigma-Aldrich | B3640 |
| Poloxamer 407 (commercial name Pluronic F-127) | Sigma-Aldrich | P2443-250G |
| Biotinylated tubulin | Cytoskeleton, Inc. | T333P-A |
| Stabilized microtubules | Prepared in house | Prepared in house |
| TetraSpeck beads | ThermoFisher Scientific | T7279 |

ACKNOWLEDGMENTS:

We thank Yin-Wei Kuo for preparing the purified eGFP-kinesin used in this study.

DISCLOSURES:

The authors have no conflicts of interest to disclose.